\documentclass[10pt,prl,twocolumn,showpacs,amsmath,amssymb]{revtex4}
\usepackage{graphicx}% Include figure files
\usepackage{dcolumn}% Align table columns on decimal point
\usepackage{bm}% bold math
\graphicspath{{Figures/}}

\begin{document}

\title{Shape transition and propulsive force of an elastic rod rotating in a viscous fluid}
%\title{The Instability of a Tilted Elastic Filament Rotating in a Viscous Fluid}

\author{Bian Qian}
\author{Thomas R. Powers}%\email{thomas_powers@brown.edu}
\author{Kenneth S. Breuer}
\email{kbreuer@brown.edu} %delete? not necessary these days?
\affiliation{Division of Engineering, Box D, Brown University, Providence, RI
02912, USA}
\date{\today}
\begin{abstract}

The deformation of thin rods in a viscous liquid is central to the mechanics of motility in cells ranging from \textit{Escherichia
coli} to sperm. Here we use experiments and theory to study the shape transition of a flexible rod rotating in a viscous fluid driven either by constant torque or at constant speed.  The rod is tilted relative to the rotation axis. At low applied torque, the rod bends gently and generates small propulsive force. At a critical torque, the rotation speed increases abruptly and the rod forms a helical shape with much greater propulsive force. We find good agreement between theory and experiment.

\end{abstract}

\pacs{ 87.16.Qp,  47.15.G-, 46.70.Hg}
%87.16.Qp Pseudopods, lamellipods, cilia, and flagella
%47.15.G- Low-Reynolds-number (creeping) flows
%46.70.Hg Membranes, rods, and strings

%\keywords{}

\maketitle

% text

Understanding how flagella and cilia work is a central aim of the field of cell motility. The problem may be split into two parts,
both of which involve physics: the means of actuation, and the fluid-structure interaction. In this letter we consider the fluid-structure interaction for thin filaments in a viscous fluid. At micron scales, viscous effects dominate inertia, and the
fluid-structure interaction problem simplifies because the Stokes equations governing the fluid motion are linear. Gray and Hancock used this linearity to develop a simple theory that successfully predicted the swimming speed of a sperm cell with a load-independent pattern of bending waves propagating along the flagellum~\cite{gray_hancock1955}.  Soon after, Machin considered
the  fluid-structure interaction~\cite{machin1958}. He argued that the motors  must be distributed all along the length of the
flagellum, since, for small amplitudes, a passive flexible rod waved at one end has an exponentially decaying envelop of deflection, whereas the amplitude of deflection in real flagellar bending waves increases slightly with distance from the head~\cite{machin1958}. The shapes and propulsive forces  of a passive rod actuated at one end have recently been examined
theoretically~\cite{wiggins_goldstein1998,YuLaugaHosoi2006} and experimentally~\cite{YuLaugaHosoi2006}. Although sperm flagella
are not passive, the results of~\cite{machin1958,wiggins_goldstein1998,YuLaugaHosoi2006} are important for modeling real flagella since the modes that Machin found also enter models in which the flagellum is actuated along its entire length~\cite{camalet_et_al1999}.

\begin{figure}
    \includegraphics[width=2.8in]{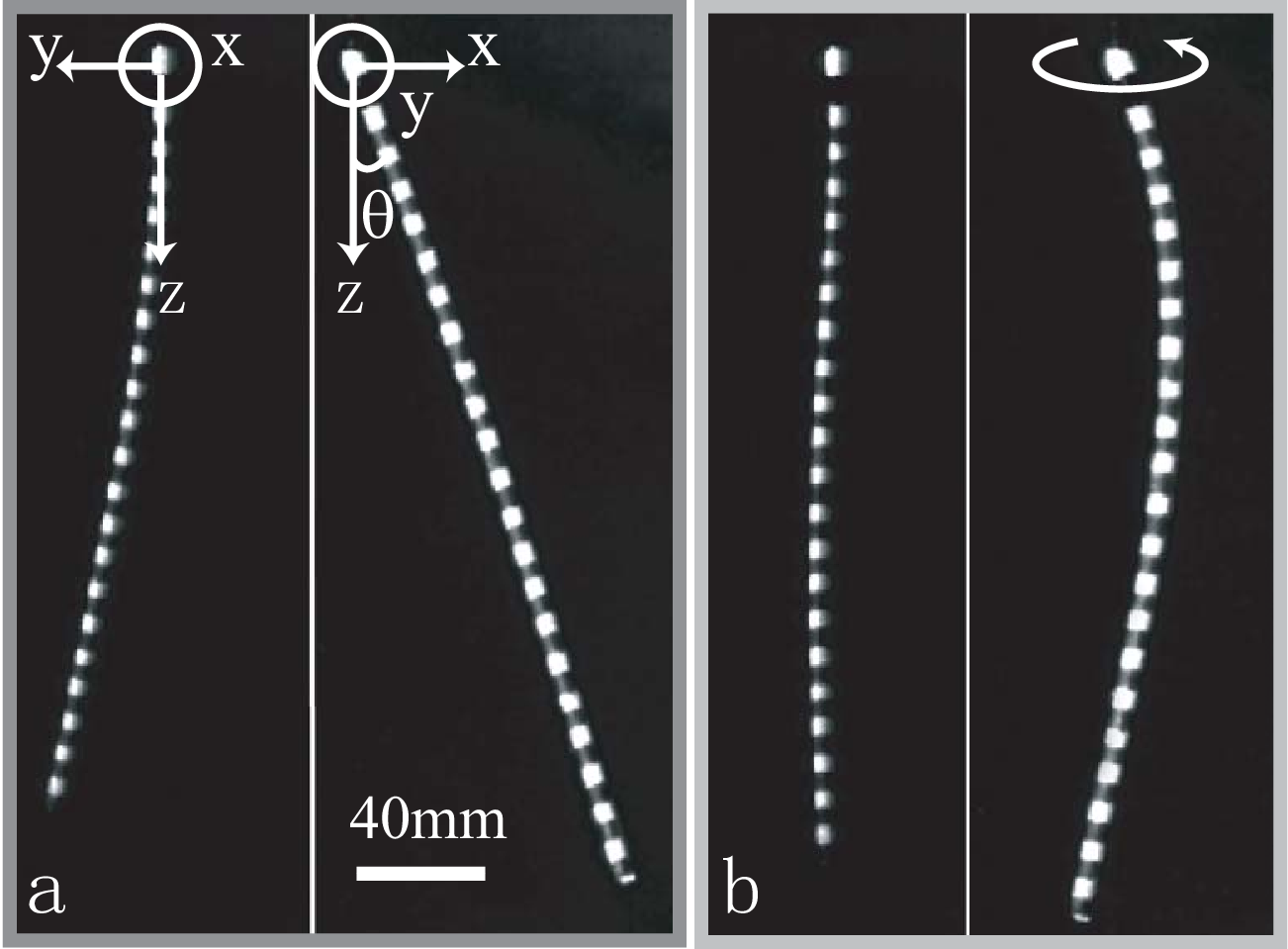}
    \caption{Orthogonal images of steady-state shapes of
    rotating rod with torque just below (a) and just above (b) the critical torque.
    The motor (not shown) is at the top, with rotation axis along $z$. Gravity points down.
    In (a) and (b), the left panel is the side view, and the right panel is the front view.
    The rod is marked with white dots for contrast.
    The axes in (b) are the same as in (a). The curved arrow in (b) denotes the sense of rotation of the rod. 
    } \label{fig_rodexp} %
\end{figure}

Rotating flagella are also common. For example, nodal cilia~\cite{PraetoriusSpring2005} have an internal structure similar to that of sperm flagella. However, instead of beating in a plane like most sperm flagella, nodal cilia rotate along the surface of an imaginary cone. The flow set up by these flagella has been implicated in the formation of left-right asymmetry in developing embryos~(see \cite{PraetoriusSpring2005} and references therein). Bacterial flagella provide another example. These flagella are helical, much thinner than eukaryotic flagella, and driven by a rotary motor embedded in the cell wall. Fluid-structure interactions are important for polymorphic transformations in swimming bacteria~\cite{turner_ryu_berg2000} and the bundling of multiple flagella~\cite{kim_etal2003}.

Complementary to the problem of understanding how biological flagella work is the problem of building an artificial microscopic
flagellum-propelled swimmer. This feat was recently accomplished by Dreyfus et al., who used a rotating external magnetic field to generate propagating planar bending waves in a filament composed of a string of colloidal magnetic particles~\cite{Dreyfus_etal2005}. A major challenge in building an artificial microscopic swimmer is the means of actuation. Manghi et al. proposed a simple mechanism in which a microscopic flexible rod rotates along the surface of an imaginary cone~\cite{ManghiSchlagbergerNetz2006,ManghiSchlagbergerKimNetz2006}. Using numerical methods, they predicted that at a critical driving torque the rod will undergo a discontinuous transition to a helical shape with significant propulsive force, independent of
the sense of rotation.  In this letter, we present a macroscopic experimental realization of this system. We also present new
theoretical results that complement the hydrodynamic calculations presented in~\cite{ManghiSchlagbergerNetz2006,ManghiSchlagbergerKimNetz2006}.

In our experiment, a servo motor rotates a flexible rod in highly viscous silicone oil. The  rod is connected to the motor shaft
with a kink such that the base of the rod  makes a fixed angle with the rotation axis (Fig.~\ref{fig_rodexp}a). The motor may be
operated either at constant speed, or at constant torque using a torque sensor with feedback. The range of torques explored was 0.5 to 8\,mN-m, and the maximum rotation frequency was less than 0.3\,Hz. The rod is a steel extension spring wrapped in Teflon$\texttrademark$ tape; the tape stiffens the rod to minimize sagging at low speed. The diameter of the rod is $a=$2.5\,mm, and the bending modulus is $A\approx3\times10^{-3}$\,N/m$^{2}$. Various rod lengths $L$ from 210\,mm to 290\,mm were tested. The silicone oil has  viscosity $\eta\approx 110$\,N-m$^2$/s and is held in a tank 420 mm on each side. With these parameters, the Reynolds number $\mathrm{Re}=\rho v L/\eta\approx(10^{-2}$, where $\rho\approx10^{3}$\,kg/m$^3$ is the fluid density, $v\approx10^{-1}$\,m/s is the typical velocity of the free end of the rod, and $L \approx10^{-1}$\,m. Front and side images of the steady-state three-dimensional shape of the rotating rod at each torque was captured using a single camera and a single mirror.  The imaging system was carefully calibrated to account for perspective, achieving an accuracy of \(\pm \)2\,mm.

\begin{figure}
    \includegraphics[height=1.25in]{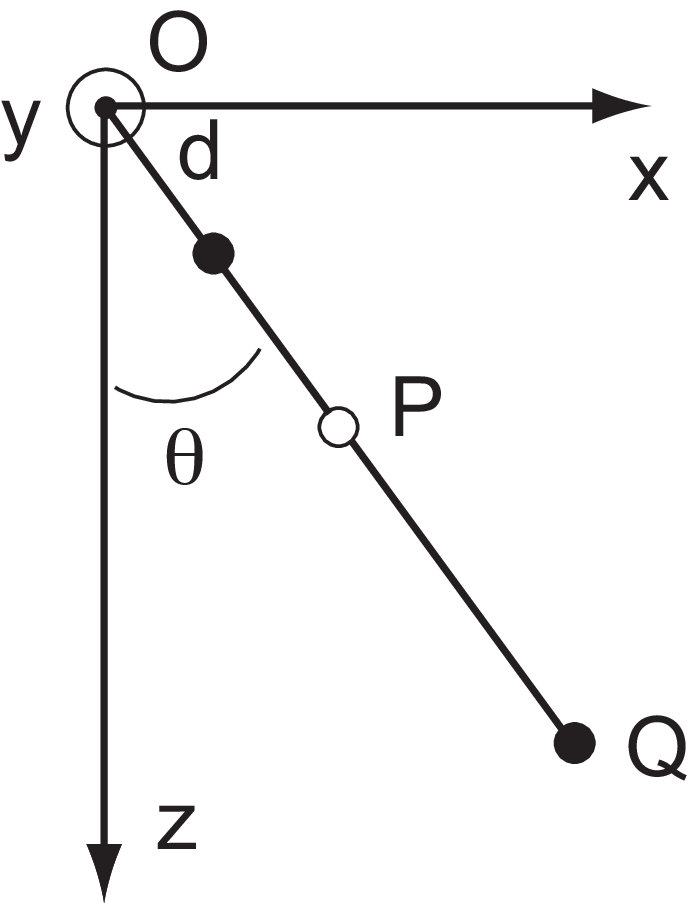}
    \caption{ Lumped parameter model consisting of two rigid links connected by a torsional spring (open circle). The top link is clamped. All drag is concentrated at the two filled circles.
    } \label{fig_ball} %
\end{figure}

\begin{figure}
  \includegraphics[width=3.75in]{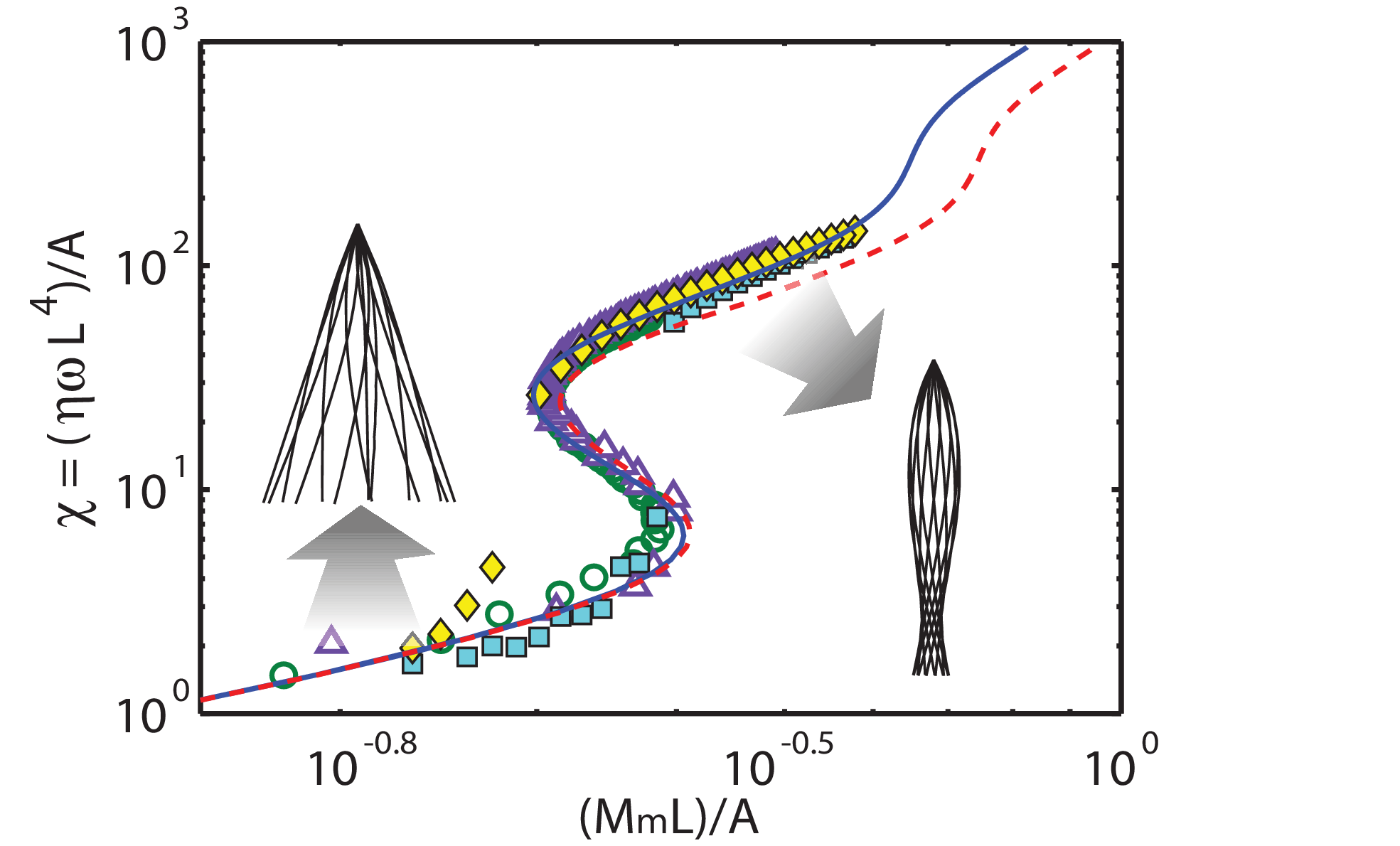}
  \caption{(Color online) Dimensionless motor torque $M_\mathrm{m}L/A$ was measured as a function of dimensionless speed $\chi$ for
  $L=250$\,mm \((\bigcirc)\) and $L=290$\,mm \((\bigtriangleup)\) with  angle $\theta=26^\circ$.
  For $L=250$\,mm, speed was measured as a function of increasing torque \((\blacksquare)\)
  and decreasing torque \((\blacklozenge)\). Note the hysteresis. The linear \((--)\) and
  nonlinear \((-)\) predictions are shown. The insets show examples of the steady state filament shapes in the low (left) and high (right) speed regimes. }\label{fig_freq_tor} %
\end{figure}

At low torque, the rod bends slightly and the rotation speed is relatively slow (Fig.~\ref{fig_rodexp}a). Just above a critical
torque, the rod adopts a helical shape and rotates much faster (Fig.~\ref{fig_rodexp}b). We first give a simple analysis of this shape transition using the lumped parameter model shown in Fig.~\ref{fig_ball}.  It is convenient to imagine prescribing motor speed $\omega$ rather
than motor torque. Since the Reynolds number is small, we take Re=0. Thus, we may work in the rod's rotating rest frame without introducing fictitious forces. The flow in this frame at point $\mathbf{r}$ is $\omega\mathbf{\hat z}\times\mathbf{r}$. The rod is modeled by two rigid links of unit length connected by a torsional spring. The link OP is fixed and represents the  condition of fixed angle $\theta$ between the rotation axis $\mathbf{\hat z}$ and the base of the rod in our experiment.
% (Fig.~\ref{fig_ball}). 
The torsional spring represents bending resistance and is only sensitive to changes in the angle between the vectors OP and PQ. Assuming $\theta\ll1$, the moment about P on PQ from the spring is $M_\mathrm{b}\approx K (\mathrm{OP}\times\mathrm{PQ})\approx K(y,2\theta-x,0)$, where $K$ is the torsional spring constant, and $(x,y,2)\approx\mathbf{r}_\mathrm{Q}$ is the position of the point Q to  leading order in $\theta$. To find the steady-state position of Q, equate the moment on PQ due to the torsional spring to the moment on PQ due to the flow. Assuming all drag on PQ is concentrated at Q (Fig.~\ref{fig_ball}), the viscous moment about P is $ M_\mathrm{v}\approx-\zeta\omega(x,y,0)$, where $\zeta$ is a resistance coefficient. Solving moment balance for $x$ and $y$ yields
$x=2\theta/[1+(\zeta\omega/K)^2]$ and $y=x \zeta\omega/K$. As $\omega$ increases from zero, the link PQ deflects and $y$ increases, which causes Q to experience a viscous force in the negative $x$ direction. These forces push Q toward the rotation axis, and tend to cause the rod in our experiment to wrap around the $z$ axis. As $\omega$ increases further, Q moves close to the rotation axis, and $y$ begins to decrease. There is also some drag on the link OP, concentrated a distance $d$ from O. The moment about O due to flow is
\begin{equation}
M_\mathrm{O}=\zeta\omega\theta^2\left( d^2+\frac{4}{1+\zeta^2\omega^2/K^2}\right).
\end{equation}
For $d^2<1/2$, we find that the moment first increases with $\omega$, then decreases as the link folds in toward the rotation axis where the flow  is slow, and then increases again as the drag from the base link OP dominates. If $M_\mathrm{O}$ is plotted vs. $\omega$, then we find an S-shaped curve, just as in our experiment (Fig.~\ref{fig_freq_tor}), with discontinuous transitions in shape and speed as moment varies.

%At low speeds, the flow in the rod frame bends the rod away from the $x$-$z$ plane (Fig.~\ref{fig_rodexp}a, left panel); $M_\mathrm{m}$ is linear in $\omega$ since the deflection is small and the Stokes equations are linear. Note that since the flow increases with distance from the $z$-axis, the hydrodynamic drag on the rod is concentrated near the free end of the rod. As $\omega$ increases, the $y$-deflection of the rod increases, and the free end
%experiences a greater flow in the negative $x$ direction, which causes the free end of the rod to wrap around the $z$-axis. By this means, the
%rod begins to assume a helical shape. As $\omega$ continues to rise, the wrapping increases, pushing the free end toward the $z$ axis, where the flow speed is lower. Once this regime is reached, $M_\mathrm{m}$ \emph{decreases} with increasing $\omega$, since the drag is still concentrated near the free end, but the speed has diminished. As $\omega$ increases further, the hydrodynamic drag shifts towards the base of the rod, which always experiences some flow due to the proscribed tilt angle $\theta$.  Hence, $M_\mathrm{m}$ once again increases with $\omega$. Figure~\ref{fig_freq_tor} shows the non-monotonic dependence of $M_\mathrm{m}$ on dimensionless rotation speed $\chi$. Note that, if $M_\mathrm{m}$ is varied instead of the speed, $\chi$, then there is a discontinuous transition in $\chi$ at the bend in the S-curve.

We now turn to a quantitative analysis of the experiment. We will continue to prescribe $\omega$ rather than motor torque $M_\mathrm{m}$
in our calculation, and limit the analysis to steady-state shapes. Unlike Manghi et al. ~\cite{ManghiSchlagbergerNetz2006,ManghiSchlagbergerKimNetz2006}, we disregard hydrodynamic interactions between distant parts of the rod and use resistive force theory to model the force per unit length $\mathbf{f}$ acting on the rod~\citep{gray_hancock1955,keller_rubinow1976}:
  \begin{equation}
     \mathbf{f}=\zeta_\perp (\mathbf{v}-\mathbf{r}_s \mathbf{r}_s\cdot \mathbf{v})
     +\zeta_\parallel \mathbf{r}_s \mathbf{r}_s\cdot \mathbf{v}  ,
     \label{slender}
  \end{equation}
where $\zeta_\perp=4\pi\eta/[\log(L/a)+1/2]$ and $\zeta_\parallel=2\pi\eta/[\log(L/a)-1/2]\approx\zeta_\perp/2$,
$\mathbf{r}(s,t)$ is the position of the point on the rod centerline with arclength $s$ at time $t$, $\mathbf{v}(s,t)$ is
the velocity of the undisturbed flow relative to the velocity of the rod at $s$, and $\mathbf{r}_s=\partial \mathbf{r}/\partial s$
is the tangent vector to the rod centerline. There is also a hydrodynamic torque per unit length distributed along the rod that
tends to twist it~\citep{wolgemuth_et_al2000a,LimPeskin2004,WadaNetz2006}. However, the effects of this torque are smaller by a factor of $(a/L)^2$ relative to effects due to translation of the rod~\cite{wolgemuth_et_al2000a} and will henceforth be disregarded. The constitutive relation for the elastic rod is
   \begin{equation}
      \mathbf{M}=A\mathbf{r}_s\times\mathbf{r}_{ss},
      \label{constitutive}
   \end{equation}
where $\mathbf{M}$ is the moment due to internal stresses exerted
on the cross-section of the rod at $s$, and $A$ is the bending
modulus~\cite{landau_lifshitz_elas}. The shape of the rod is
determined by force and moment balance,
   \begin{eqnarray}
     \mathbf{F}_s+\mathbf{f}+\mathbf{f}_\mathrm{g}&=&\mathbf{0} \label{force_bal}\\
     \mathbf{M}_s+ \mathbf{r}_s\times\mathbf{ F}&=&\mathbf{0}, \label{moment_bal}
   \end{eqnarray}
where $\mathbf{F}(s)$ is the force due to internal stresses acting on the rod cross section at $s$, and $\mathbf{f}_\mathrm{g}=(\mu_\mathrm{rod}-\pi a^2\rho_\mathrm{oil})g\mathbf{\hat z}$ is the buoyancy force per unit length due to the density difference between the rod (linear density $\mu_\mathrm{rod}=0.0478$\,kg$/$m with oil inside) and silicone oil ($\rho_\mathrm{oil}=970$\,kg$/$m$^3$). The boundary conditions are $\mathbf{r}(0)=\mathbf{0}$, $\mathbf{r}_s(0)=\mathbf{\hat x}\sin\theta+\mathbf{\hat z}\cos\theta$, $\mathbf{F}(L)=\mathbf{0}$, and $\mathbf{M}(L)=\mathbf{0}$ \cite{landau_lifshitz_elas}. As in the lumped parameter model, $\mathbf{v}=\omega\mathbf{\hat z}\times\mathbf{r}$ at steady state in the rod frame.

The primary dimensionless groups governing the rod shape are the  angle $\theta$ and the dimensionless rotation speed $\chi=\eta\omega L^4/A=(L/\ell)^4$, where $\ell=[A/(\eta\omega)]^{1/4}$ is the characteristic length scale determined by bending resistance and viscous drag~\cite{machin1958,wiggins_goldstein1998}. In addition, the aspect ratio $L/a$ and the non-dimensional gravitational force $g\mu_{\mathrm{rod}}L^3/A$ are included in the analysis, but not parametrically explored since they play a minor role. Figure~\ref{fig_freq_tor} shows that experimental measurements using two rod lengths collapse well onto a single curve when dimensionless speed, $\chi$, is plotted against dimensionless motor
torque, $M_\mathrm{m}L/A$. The open symbols represent constant velocity rotation and trace out the entire S-shaped curve.  The filament shape is stable at every prescribed value of $\chi$, and the shape changes continuously from the slightly bent shape to the helical shape as $\chi$ increases. The closed symbols represent constant torque operation. There is a jump in rotation speed and filament shape at two different torque values, depending if torque is ascending or descending. For descending torque, the time to reach steady state is prohibitively long, and the diamond symbols above the curve in Fig.~\ref{fig_freq_tor} represent shapes that are still relaxing slowly to steady state.

\begin{figure}
  \includegraphics[width=3.25in]{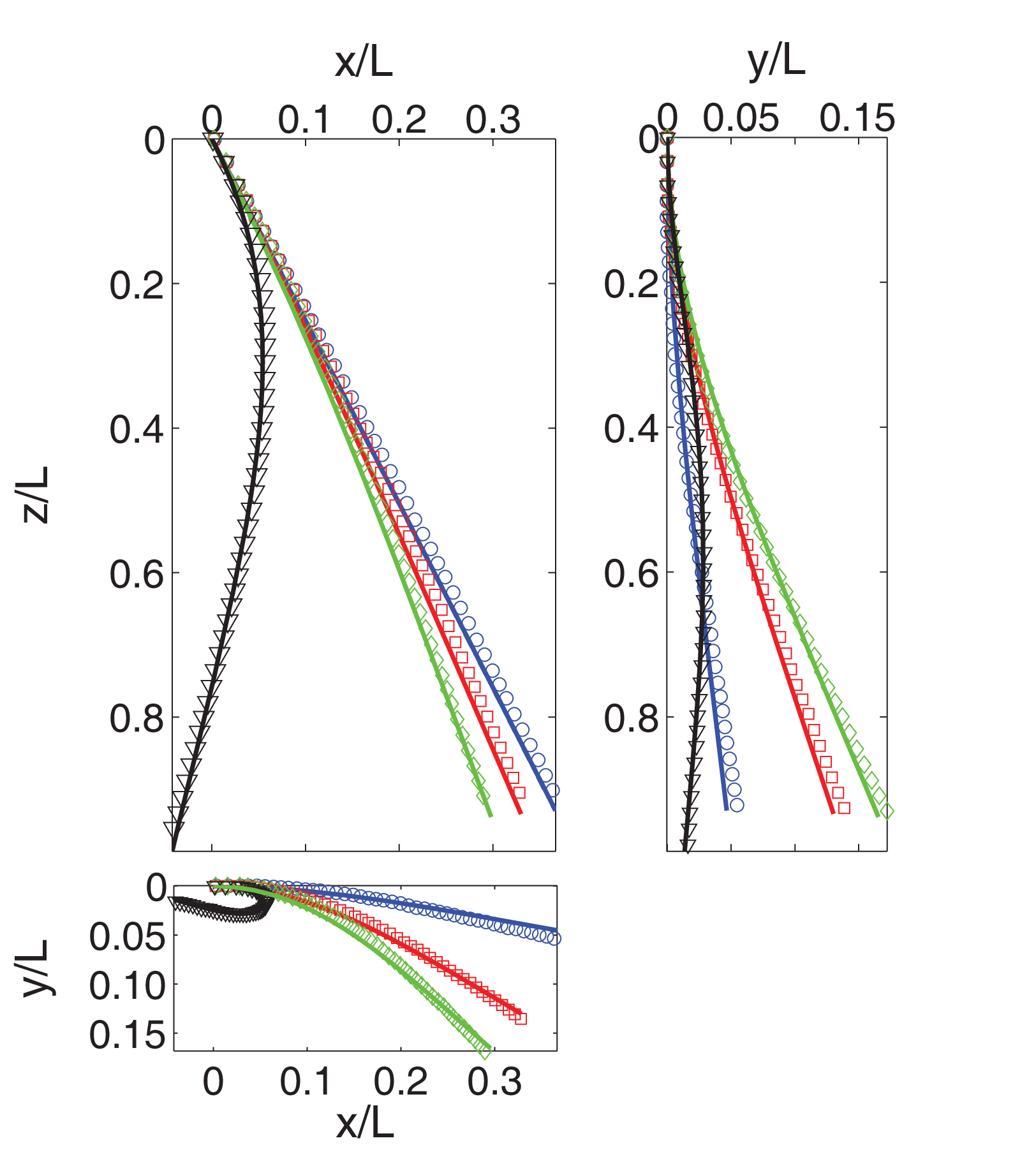}
  \caption{\label{fig:epsart}(Color online) Steady-state shapes of a rotating rod
  from experimental measurements for \(\chi= 1.38\) \((\bigcirc)\),
  4.25 \((\square)\), 5.91 \((\lozenge)\) (before transition), and
  164.63 \((\bigtriangledown)\) (after transition), along with the shapes
  calculated from the nonlinear \((-)\) theory. The rod has \(L=210\)\,mm and \(\theta=20^\circ\). 
  } \label{fig_projection} %
\end{figure}

It is interesting to note that the nonlinear behavior of the speed-torque curve displayed in Fig~\ref{fig_freq_tor} can be
qualitatively explained using linear approximations valid for small rod deflections, since the shape of the rod depends
nonlinearly on $\chi$ even when the rod deflection is small. For small $\theta$, the rod is aligned mainly along the $z$-axis and, disregarding gravity, the deflection $\mathbf{r}_\perp(z)=(x(z),y(z))$ obeys
\begin{equation}
  %-A \frac{\partial^4 \mathbf{r}_\perp}{\partial z^4}+\zeta_\perp\omega\mathbf{\hat z}\times\mathbf{r}_\perp=\mathbf{0}.
  -\ell^4 \frac{\partial^4 \mathbf{r}_\perp}{\partial z^4}+\mathbf{\hat z}\times\mathbf{r}_\perp=\mathbf{0}.
  \label{lineom}
\end{equation}
The solution to Eq.~(\ref{lineom})  is a generalization of Machin's solution to the in-plane bending problem~\cite{machin1958}, and of the same form as the solution for a flexible rod held parallel to but some distance from the axis of rotation~\cite{powers2002}.

To calculate the motor torque $M_\mathrm{m}$ required to rotate the rod at speed $\omega$, observe that the moment due to viscous drag must equal the elastic moment at the base of the rod:
\begin{equation}
  M_\mathrm{m}=-\mathbf{\hat z}\cdot\int\mathbf{r}\times\mathbf{f}\mathrm{d}s
  =-A\mathbf{\hat z}\cdot\mathbf{r}_s\times\mathbf{r}_{ss}(0).
  \label{moments}
\end{equation}
The second equality of (\ref{moments}) follows from (\ref{force_bal}--\ref{moment_bal}). The results of the linear calculation for driving torque vs. speed are shown in Fig.~\ref{fig_freq_tor} along with the experimental data for $\theta=26^\circ$. For small $\chi$, $M_\mathrm{m}$ increases linearly with $\chi$. We may use a simple scaling argument to capture the dependence of $M_\mathrm{m}$ on $\chi$ for $\chi\gg1$. For large $\chi$, Eq.~(\ref{lineom}) implies that the shape of the rod is helical
with an envelope that decays exponentially with length scale $\ell$. Assuming isotropy $x\sim y$ and using force balance (\ref{lineom}), the viscous force per length $f\sim\eta\omega y\sim Ay/\ell^4$, which  implies a total viscous moment $M_\mathrm{v}L/A\sim Ly^2/\ell^3$. On the other hand, the bending moment at the base of the rod scales as $M_\mathrm{b}L/A\sim yL/\ell^2$. Equating $M_\mathrm{v}$ and $M_\mathrm{b}$ yields $y\sim \ell=L\chi^{-1/4}$~\cite{koehler_powers2000}. Thus, for $\chi\gg1$, the motor torque must scale as $M_\mathrm{m}L/A\sim\chi^{1/4}$. 
%This scaling is also obeyed for $\chi\gg1$ by the solution to the linear theory shown in Fig.~\ref{fig_freq_tor}. 
Unfortunately, our experiment cannot access this high-speed regime due to limitations in the torque-speed characteristic of our motor. We also observe an intermediate scaling, $M_\mathrm{m}L/A\sim\chi^{1/2}$. This scaling arises since in this sub-asymptotic regime the deflection $y\sim L\chi^{-1/4}$, but the scale for bending of the rod is still $L$ and not $\ell$ .

For large $\theta$, the deflection of the rod is significant even for small $\chi$, and the linear theory is inaccurate. However,
the nonlinear equations~(\ref{slender}--\ref{moment_bal}) are readily solved with shooting methods~\cite{PressVetterlingTeukolsky1992}. The nonlinear theory gives a more accurate prediction for the speed-torque relationship in the high-speed regime where the linear and nonlinear theories differ (Fig.~\ref{fig_freq_tor}). As $\theta$ increases, the general appearance of the torque-speed relationship remains unchanged although both the critical torque and the jump in speed at the transition increase (not shown here). Finally, for large $\chi$, the scaling analysis presented above remains valid, and the moment scales like $\chi^{1/4}$ for $\chi\gg1$, just as in the linear theory.

Figure~\ref{fig_projection} shows the steady-state rod shapes for four different values of $\chi=(L/\ell)^4$, comparing experimental data (symbols) and the nonlinear theory (solid line). The agreement between theory and experiment is good. Note that the $y$-$z$ projection shows how $y(L)$ first increases with $\chi$ and then decreases, in accord with our intuitive argument.

\begin{figure}
  \includegraphics[width=3.2in]{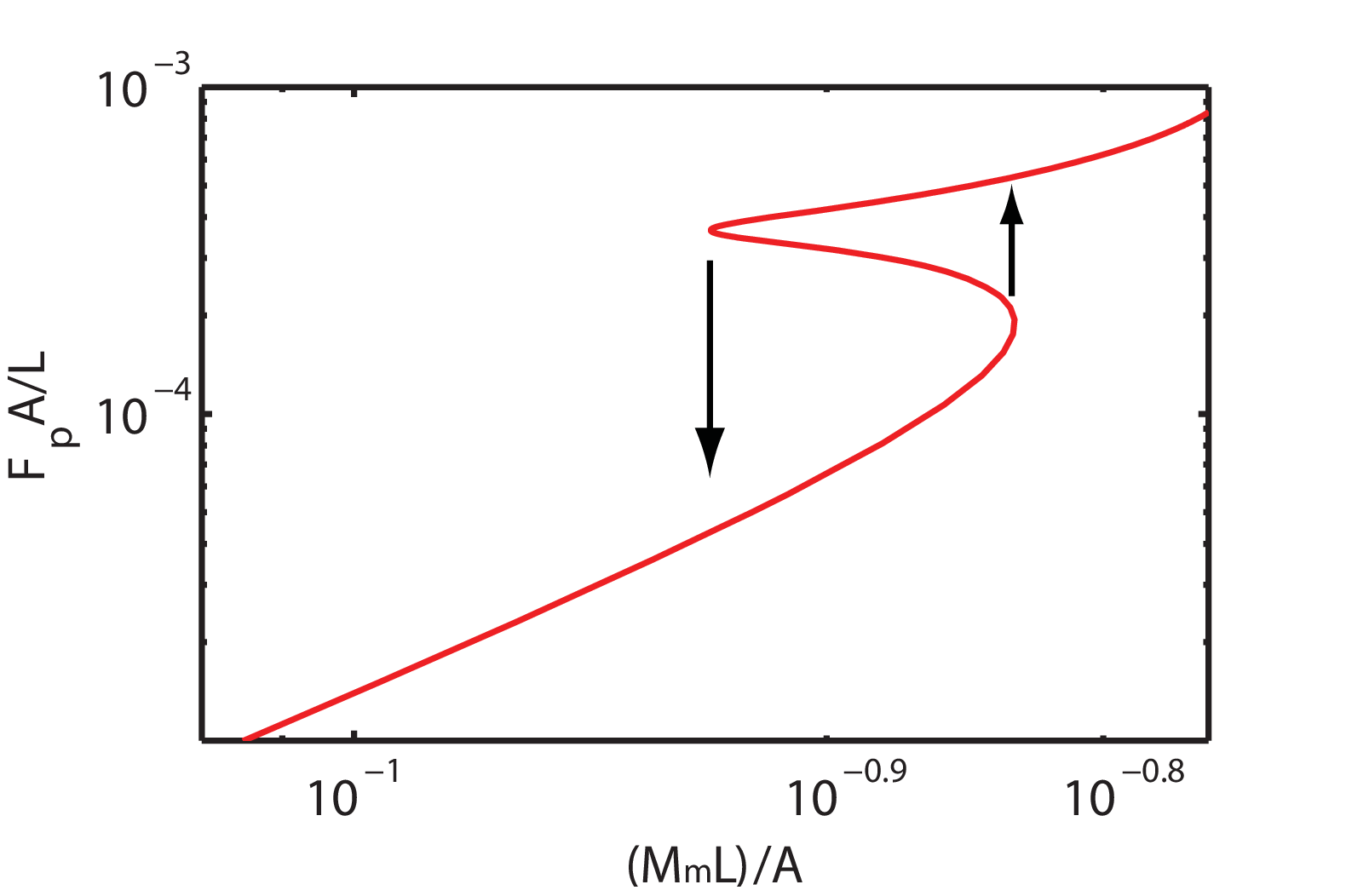}
  \caption{(Color online) Theoretical calculation of the dimensionless propulsion
  force $F_\mathrm{p}$ as a function of dimensionless \(M_\mathrm{m}\) for \(\theta = 20^\circ\),
  using the nonlinear equations for the ideal case of zero
  gravity. The arrows denote the transition for ascending \((\uparrow)\) and descending torque \((\downarrow)\). 
  }\label{fig_prof} %
\end{figure}

We calculate the thrust, or axial force from the rod on the motor, from the shape of the rod using $F_\mathrm{p}=\mathbf{\hat z}\cdot\int\mathbf{f}\mathrm{d}s=-\mathbf{\hat z}\cdot\mathbf{F}(0)$. The kinematic reversibility of Stokes flow implies that a rigid
rod rotating along the surface of a cone generates zero propulsive thrust. For small $\chi$, the elastic rod deforms slightly and generates little thrust. Above the critical torque, as the helical shape develops, the thrust increases abruptly (Fig.~\ref{fig_prof}). Since the shape of an actuated elastic filament cannot be decoupled from swimming kinematics~\cite{LaugaFloppy2007}, it would be an interesting generalization of our work to build an artificial swimmer driven by a rotating elastic rod, tilted at the
base to the rotation axis.

This work was supported in part by National Science Foundation Grants Nos. CTS-0508349 (KSB) and NIRT-0404031 (TRP). As we were completing this letter, we learned of the work of the group of
Fermigier, who have conducted a similar experiment~\cite{Fermigier_etal}.

\bibliographystyle{apsrev}
\bibliography{therefs}

\end{document}